\documentclass[twocolumn,preprintnumbers,amsmath,amssymb,superscriptaddress,nofootinbib]{revtex4}
\usepackage{graphicx}

\begin{document}
\title{Gravitational waves generated from the cosmological QCD phase transition within AdS/QCD}
%\author{M. Ahmadvand and K. Bitaghsir Fadafan}
%\email{}
\author{M. Ahmadvand}
\affiliation{Physics Department, Shahrood University of Technology, P.O.Box 3619995161 Shahrood, Iran}
\affiliation{School of Physics, Institute for Research in Fundamental Sciences (IPM), Tehran, Iran}
\author{K. Bitaghsir Fadafan}
\affiliation{Physics Department, Shahrood University of Technology, P.O.Box 3619995161 Shahrood, Iran}
\date{\today}

\begin{abstract}
We study the gravitational waves produced by the collision of the bubbles as a probe for the cosmological first order QCD phase transition, considering heavy static quarks. Using AdS/QCD and the correspondence between a first order Hawking-Page phase transition and confinement-deconfinement phase transition, we find the spectrum and the strain amplitude of the gravitational wave within the hard and soft wall models. We postulate the duration of the phase transition corresponds to the evaporation time of the black hole in the five dimensional dual gravity space, and thereby obtain a bound on the string length in the space and correspondingly on the duration of the QCD phase transition. We  also show that IPTA and SKA detectors will be able to detect these gravitational waves, which can be an evidence for the first order deconfinement transition.
\end{abstract}
\maketitle

\section{Introduction}
According to the standard cosmology, during the evolution of the universe, several Phase Transitions (PTs) have occurred. One of the consequences of these out of equilibrium events is the generation of the Gravitational Waves (GWs), which open a new way to explore the early universe, during and after PTs. However, if the cosmological PT is not a first order type, it cannot give rise to the GW production. For a cosmological first order PT, there are two degenerate thermodynamical states separated through a barrier. Tunneling to the new phase is proceeded by the bubble nucleation. The expansion of these bubbles and finally their collision with each other result in the production of GWs \cite{Jinno:2015doa, Caprini:2015zlo}. Furthermore, other different sources including inflationary quantum fluctuations \cite{Ashoorioon:2013oha}, cosmic strings \cite{Rubin:2001yw}, and Black Hole (BH) collisions \cite{East:2014nfa} for radiating GWs have been proposed. Therefore, GWs may be applied as a powerful probe on astrophysical and cosmological events in the universe.\\Numerical calculation indicates QCD PT at finite temperature for small and large quark masses was a first order PT \cite{Philipsen:2010gj}. However, as lattice QCD shows, the transition is an analytic crossover for intermediate quark masses, particularly for three physical light quarks and small chemical potential \cite{Bhattacharya:2014ara}.\\ Physical QCD has approximate chiral and Z(3) center symmetries for very small and large quark masses, respectively. Three light pions and color screening are remainders of these spontaneously broken symmetries. For intermediate masses these symmetries are explicitly broken and related order parameters are non-zero at all temperatures.\\The expectation value of Polyakov loop which can be read off from the heavy quark potential is the relevant order parameter for the confinement-deconfinement PT. In this work, we focus on the cosmological QCD PT. At temperatures that the deconfinement transition occurred, heavy quarks lost their dynamical importance. Thus we consider pure gauge theory with non-dynamical heavy quarks appropriate for Z(3) symmetry broken in the confinement-deconfinement PT with the expectation value of Polyakov loop as the order parameter. Employing the point that the confinement-deconfinement PT is corresponded to Hawking-Page PT \cite{haw} which is of a first order, we explain this PT in the AdS/QCD context and explore it through possible GWs generated during the transition.\footnote{different scenarios including a short inflation during the QCD PT \cite{boe} and a model with a large neutrino chemical potential \cite{Schwarz:2009ii} have been suggested to provide a first order QCD PT (besides, see \cite{Schettler:2010dp})} \\After the conjecture of AdS/CFT correspondence and its generalization to gauge/gravity, people have attempted to get a better understanding of QCD and its properties by using the dual gravity theory in five dimensions; the interpretation of the Hawking-Page PT as the confinement-deconfinement PT \cite{Witten:1998zw} and the computation of the energy loss of heavy quarks moving in the quark gluon plasma \cite{Gubser:2006bz} are investigations to fulfill this goal.\\The precise dual gravity describing the real QCD is not yet found. However, so-called AdS/QCD is a promising extension which can explain important features of QCD. The AdS/QCD top-down approach first takes into account a string theory, then deforms the dual super Yang-Mills theory to gain QCD properties such as confinement \cite{Polchinski:2000uf} while in the other approach, bottom-up, starting with QCD, the dual description is constrained by QCD ingredients. The conformal isometry of AdS space is consistent with UV asymptotic freedom of QCD. In the hard wall model of this approach \cite{Polchinski:2001tt, Erlich:2005qh} to produce confinement, the small radius region of AdS is truncated, while for the soft wall model this truncation is smoothly performed by a dilaton field \cite{Karch:2006pv}. In \cite{Herzog:2006ra}, the author finds the confinement-deconfinement PT for these two models and the PT temperature obtained in the soft wall model is very close to the prediction of a lattice calculation.\\Using holographic description of the hard and soft wall models, we try to estimate GWs generated from confinement-deconfinement PT. We relate the duration of the PT to the evaporation time of the BH in these AdS spaces and attain a bound on the string length. The generated GWs are obtained for $ N=3 $ (where $ N $ is the number of colors); extrapolation to large $ N $ leads to the stronger PT. Moreover, we display these GWs can be detected by International Pulsar Timing Array (IPTA) \cite{ipta} and Square Kilometre Array (SKA) \cite{ska} detectors.\\ Section II is devoted to the first order PT parameters characterizing the GW. In section III, we introduce the gravity setups and calculate GWs in the hard and soft wall models. We finally summarize the outcomes.
\section{gravitational waves and the first order cosmological qcd phase transition}
Based on Einstein's prediction of GWs, space-time fluctuation propagates as a wave at the speed of light from their sources. In general relativity, these GWs can be set by linearizing Einstein equation of motion (e.o.m). As pointed before out, there are various sources for the GW radiation and we study the sort of the GW generated from cosmological PTs.\\If a first order PT takes place, the transition into the true vacuum proceeds due to bubble nucleation and percolation. Dynamics of these bubbles plays an important role in GW production. There are two sorts of combustion modes: detonation and deflagration. When PT wall expands faster than the speed of sound, combustion occurs through detonation and for deflagration, bubble front moves at the subsonic velocity.\\Bubbles expand and collide with each other and part of the stored energy in the walls is converted to GWs. From this process, there are three sources for the GW production: bubble collision, sound waves and Magnetohydrodynamic (MHD) turbulence after the bubbles collided. The GW contribution from the bubble collision is calculated by a numerical method \cite{Jinno:2015doa} known as envelope approximation \cite{Kamionkowski:1993fg} which simulates PTs with the envelope of bubbles and expresses GW spectrum in terms of first order PT parameters. Moreover, the GW contribution of sound waves \cite{Hindmarsh:2013xza} and MHD turbulence \cite{Caprini:2006jb}, considered as Kolmogorov-type turbulence, has been computed.\\GW properties calculated by the  numerical methods are given by first order PT parameters. Important parameters of a first order PT affecting GW properties are: the ratio of the vacuum energy density to the thermal energy density of the universe at the PT time, $ \alpha $, the duration of the PT, $ \tau ^{-1} $, the velocity of bubble expansion, $ v_b $, the temperature at which the PT occurs, $ T_* $, and the fraction of the vacuum energy which is converted to the kinetic energy of the bubbles and the fluid motion rather than reheating the fluid, $ \kappa $.\\The contribution of the GW energy density from the mentioned sources is given by (we assume the so-called runaway bubble walls with the ultra-relativistic velocity and three GW sources) \cite{Huber:2008hg, Caprini:2015zlo}
\begin{equation}\label{1}
h^2\Omega (f)=h^2\Omega _{en}(f)+h^2\Omega _{sw}(f)+h^2\Omega _{tu}(f),
\end{equation}
where
\begin{eqnarray}\label{2}
h^2\Omega _{en}(f)&=&3.5\times 10^{-5}\Big(\frac{0.11 v_b^3}{0.42+v_b^2}\Big) \Big(\frac{H_*}{\tau}\Big)^{2}\Big(\frac{\kappa \alpha}{1+\alpha}\Big)^2\nonumber \\ &\times &\Big(\frac{10}{g_*}\Big)^{\frac{1}{3}}S_{en}(f),\nonumber \\
h^2\Omega _{sw}(f)&=&5.7\times 10^{-6}\Big(\frac{H_*}{\tau}\Big)\Big(\frac{\kappa _v \alpha}{1+\alpha}\Big)^2\Big(\frac{10}{g_*}\Big)^{\frac{1}{3}} v_b~ S_{sw}(f), \nonumber \\
h^2\Omega _{tu}(f)&=&7.2\times 10^{-4}\Big(\frac{H_*}{\tau}\Big)\Big(\frac{\kappa _{tu} \alpha}{1+\alpha}\Big)^{\frac{3}{2}}\Big(\frac{10}{g_*}\Big)^{\frac{1}{3}} v_b~ S_{tu}(f).\nonumber \\
\end{eqnarray}
The spectral shapes of GWs are characterized by the numerical fits as \cite{Huber:2008hg, Caprini:2015zlo}
\begin{eqnarray}\label{3}
S_{en}(f)&=&\frac{3.8(\frac{f}{f_{en}})^{2.8}}{1+2.8(\frac{f}{f_{en}})^{3.8}},\nonumber \\
S_{sw}(f)&=&\Big(\frac{f}{f_{sw}}\Big)^3\Big(\frac{7}{4+3(\frac{f}{f_{sw}})^{2}}\Big)^{\frac{7}{2}},\nonumber \\
S_{tu}(f)&=&\frac{(\frac{f}{f_{tu}})^3}{(1+\frac{f}{f_{tu}})^{\frac{11}{3}} (1+\frac{8\pi f}{h_*})},
\end{eqnarray}
with
\begin{equation}\label{4}
h_*=1.1\times 10^{-8} [\mathrm{Hz}]\Big(\frac{T_*}{100~\mathrm{MeV}}\Big)\Big(\frac{g_*}{10}\Big)^{\frac{1}{6}}.
\end{equation}
$ f_{en, sw, tu} $ are the peak frequency of each GW spectrum given by
\begin{eqnarray}\label{5}
f_{en}=11.3\times 10^{-9} [\mathrm{Hz}] \Big(\frac{f_*}{\tau}\Big)\Big(\frac{\tau}{H_*}\Big)\Big(\frac{T_*}{100~\mathrm{MeV}}\Big)\Big(\frac{g_*}{10}\Big)^{\frac{1}{6}},\nonumber \\
f_{sw}=1.3\times 10^{-8} [\mathrm{Hz}] \Big(\frac{1}{v_b}\Big)\Big(\frac{\tau}{H_*}\Big)\Big(\frac{T_*}{100~\mathrm{MeV}}\Big)\Big(\frac{g_*}{10}\Big)^{\frac{1}{6}},\nonumber \\
f_{tu}=1.8\times 10^{-8} [\mathrm{Hz}] \Big(\frac{1}{v_b}\Big)\Big(\frac{\tau}{H_*}\Big)\Big(\frac{T_*}{100~\mathrm{MeV}}\Big)\Big(\frac{g_*}{10}\Big)^{\frac{1}{6}}
\end{eqnarray}
where
\begin{equation}\label{6}
\frac{f_*}{\tau}=\frac{0.62}{1.8-0.1 v_b+v_b^2}.
\end{equation}
$ \kappa $, $ \kappa _v $ and $ \kappa _{tu} $ parameters are the fraction of the vacuum energy converted to the kinetic energy of the bubbles, bulk fluid motion and the MHD turbulence, respectively \cite{Espinosa:2010hh, Caprini:2015zlo}:
\begin{equation}\label{7}
\kappa =1-\frac{\alpha _{\infty}}{\alpha},~~~~\kappa _v=\frac{\alpha _{\infty}}{0.73+0.083\sqrt{\alpha _{\infty}}+\alpha _{\infty}},~~~~\kappa _{tu}=\epsilon \kappa _v
\end{equation}
where $ \epsilon $ is of order $ 0.05-0.1 $ \cite{Hindmarsh:2015qta} and $ \alpha _{\infty} $ is the minimum value of $ \alpha $ due to which bubbles can run away
\begin{equation}\label{8}
\alpha_{\infty}=\frac{30}{48 \pi ^2}\frac{\sum _a N_a \Delta m_a ^2}{g_* T_*^2}.
\end{equation}
$ N_a $ denotes the number of degrees of freedom for fermion species and $ \Delta m_a $ is the mass difference of the particles between two phases (for details see \cite{Espinosa:2010hh}). We assume $ \epsilon =0.05 $ and also $ v_b =1 $.
The characteristic strain amplitude produced by GW is defined as
\begin{equation}\label{9}
h_c(f)=1.3 \times 10^{-18}\Big(\frac{1 \mathrm{Hz}}{f}\Big)\Big(h^2\Omega (f)\Big)^{\frac{1}{2}}.
\end{equation}
Moreover, other parameters are defined as follows
\begin{equation}\label{10}
\alpha =\frac{\epsilon _*}{\frac{\pi^2}{30}g_*T_*^4},
\end{equation}
and the related vacuum energy at the PT is
\begin{equation}\label{11}
\epsilon _*=\Big(-\Delta F(T)+T\frac{d \Delta F(T)}{dT}\Big)\Bigg|_{T=T_*}.
\end{equation}
The difference between free energies of two phases is denoted by $ \Delta F $ and the Hubble parameter at the temperature $ T_* $ is given by
\begin{equation}\label{12}
H_*=\sqrt{\frac{8\pi^3 g_*}{90}}\frac{T_*^2}{m_{pl}},
\end{equation}
where the number of effective relativistic degrees of freedom at the PT is $ g_*= 10 $ and the Planck mass is $ m_{pl}=1.22\times 10^{22}~\mathrm{MeV} $. To calculate QCD PT parameters, we use hard and soft wall models in the AdS/QCD correspondence context.
\section{gravity setup}
\subsection{Hard Wall}
According to Hawking-Page PT, there is a first order PT between Schwarzschild-AdS BH and thermal AdS spaces. In \cite{Witten:1998zw}, Witten argued in the dual gauge theory the Schwarzschild-AdS BH corresponds to deconfinement at the high temperature and thermal AdS corresponds to confining phase at the low temperature for the compact boundary, while for the non-compact boundary there is no PT. However, by introducing IR cut-off in Poincar$\acute{\mathrm{e}}$ AdS spaces, hard wall model, \cite{Herzog:2006ra} showed that Hawking-Page PT would be possible.\\In this section we obtain necessary parameters within this model. The five dimensional gravity action dual to gluodynamics with static heavy quarks and negligible baryon chemical potential is given by
 \begin{equation}\label{13}
S=-\frac{1}{16\pi G_5}\int d^5x \sqrt{g}~\Big(\mathcal{R}+\frac{12}{R^2}\Big),
\end{equation}
where $ \mathcal{R} $ is the Ricci scalar, $ -12/R^2  $ is the negative cosmological constant, and $ R $ is the AdS radius. Two solutions for the e.o.m are an Euclidean AdS in Poincar$\acute{\mathrm{e}}$ coordinate
\begin{equation}\label{13}
ds^2=\frac{R^2}{z^2}\Big(dt^2+d\vec{x}^2+dz^2\Big),
\end{equation}
where the radial coordinate is limited to $ 0<z\leq z_0 $ and $ 1/z_0 $ corresponds to IR cut-off in the dual gauge theory; the second solution is the AdS-BH whose metric is as follows
\begin{equation}\label{14}
ds^2=\frac{R^2}{z^2}\Big(f(z)dt^2+d\vec{x}^2+\frac{dz^2}{f(z)}\Big),
\end{equation}
where $ f(z)=1-z^4/z_h^4 $ and $ 0<z\leq \bar{z} $ with $ \bar{z}=\mathrm{min}(z_0, z_h) $. The thermal AdS case has temperature $ T=1/\beta ' $ where $ \beta ' $ is the period of the Euclidean time, while the Hawking temperature of the BH, $ T=1/(\pi z_h) $, is achieved from the near horizon metric.\\Since $ \mathcal{R}=-20/R^2 $ for these spaces, from (\ref{12}) we can obtain free energy densities as
\begin{equation}\label{15}
F^{AdS}\simeq\frac{4R^3T}{8\pi G_5}\int _0^{\beta '} dt\int _{\varepsilon}^{z_0}dz~z^{-5},
\end{equation}
\begin{equation}\label{16}
F^{BH}\simeq\frac{4R^3T}{8\pi G_5}\int _0^{\pi z_h} dt\int _{\varepsilon}^{\bar{z}}dz~z^{-5},
\end{equation}
where $ \varepsilon $ is the UV cut-off. Setting the BH into the space consistently at $ \varepsilon $ leads to $ \beta '\sqrt{f_{AdS}(\varepsilon)}=\pi z_h\sqrt{f_{BH}(\varepsilon)} $. In the limit of $ \varepsilon \rightarrow 0 $, $ \beta ' $ is expressed in terms of BH temperature. For $ z_0>z_h $, there is a PT at $ z_0^4=2z_h^4 $ as seen from the following equation
\begin{equation}\label{17}
\Delta F=\frac{R^3}{8\pi G_5}\Big (\frac{1}{z_0^4}-\frac{1}{2z_h^4}\Big ).
\end{equation}
By relating $ z_0 $ to the mass of the lightest $ \rho $ meson, one obtains $ z_0=1/(323~ \mathrm{MeV}) $ \cite{Erlich:2005qh} and thus $ T_*=122~ \mathrm{MeV} $ \cite{Herzog:2006ra}. Furthermore, we can calculate the relevant latent heat at the PT and $ \alpha $ as
\begin{equation}\label{18}
\epsilon _*=\frac{N^2\pi ^2T_*^4}{2},~~~~~~~\alpha =\frac{3N^2}{2}.
\end{equation}
We used $ G_5=8\pi ^3g_s^2\alpha _s^4/R^5 $ and $ R^4=4\pi N g_s\alpha _s^2 $ where $ g_s $ and $ \alpha _s $ are the string coupling and tension, respectively. To obtain $ \Delta F $ and $ \epsilon _* $, we assumed the UV cut-off of two spaces are the same and attained $ \beta '=\pi z_h(1-\varepsilon ^4/(2z_h^4) ) $. We also assume the transition temperature, $ T_* $, is equal to the temperature of the bubble nucleation. Moreover, we obtain $ \alpha _{\infty} $, Eq. (\ref{8}), so that $ \Delta m\approx 400 ~\mathrm{MeV} $, which is the quark mass difference between the constituent (effective) quark mass \cite{Lavelle:1995ty} and quark mass in the deconfinement phase, and $ N_a =6 $ for the  quark particles.\\As seen from Eq. (\ref{18}), the larger $ \alpha $, the stronger the PT becomes. Therefore, for the large $ N $ limit, it gives rise to the very strong PT.\\Due to the existence of a PT in these truncated AdS spaces and the notion that two phases are not stable and always thermodynamically dominated, we assume the duration of the PT, $ \tau ^{-1} $, can be found by the evaporation time, $ t_{e} $, of the BH in this space. To do so, first we should calculate the BH mass. One can get the energy density of the AdS-BH from the renormalized free energy density \cite{BallonBayona:2007vp}
\begin{equation}\label{19}
E=\frac{\partial}{\partial \beta}\beta F^{BH}=\frac{3R^3}{16\pi G_5 z_h^4}.
\end{equation}
By inspiration from a dimensional reduction carried out for 10 dimensional Newton constant to gain $ G_5 $, we expect the relevant Newton constant in this non-compact boundary, $ R^3\times S^1 $, to obtain the BH mass is $ G'_5=8\pi ^3g_s^2\alpha _s^4/(z_h^3R^2) $. Therefore, the BH mass can be attained as
\begin{equation}\label{20}
M\simeq \int _0^{z_h} dz~z_h^2\frac{3N^2}{8\pi ^2R^3z_h}\simeq \frac{N^2z_h^2}{R^3}.
\end{equation}
Then, the power of losing energy \cite{page}, $ P=A T^4\simeq z_h^{-2} $, ($ A $ is the BH surface area) gives
\begin{equation}\label{21}
\int _0^M dM~\frac{R^3M}{N^2}\simeq \int _{t_e}^0 dt.
\end{equation}
Hence, according to our assumption $ \tau ^{-1}=t_e\simeq N^2/(R^3T^4) $. Also, one may find that $ N=3 $ is consistent with the strongly-coupled SU(3) gauge theory by considering the effective string tension relation, $ \sigma =R^2/(2\pi \alpha _s z_0^2) $ \cite{BoschiFilho:2005mw}, whose value can be obtained from a quark antiquark potential energy calculation \cite{Cheng:2006qk}, $ \sqrt{\sigma}\simeq 465 $ MeV. From this result and Eq. (\ref{12}), we find that a string length of the order of $ l_s\sim\sqrt{\alpha _s} \sim 1/(10^{10}~\mathrm{MeV})-1/(10^{9}~\mathrm{MeV}) $ is corresponded to $ \tau =H_* $ and $ \tau =10H_* $, respectively.\\Now, putting the related parameters in Eqs. (\ref{1}), (\ref{2}), and (\ref{9}), we can identify the generated GW.
\begin{figure}[th]
\includegraphics[scale=0.92]{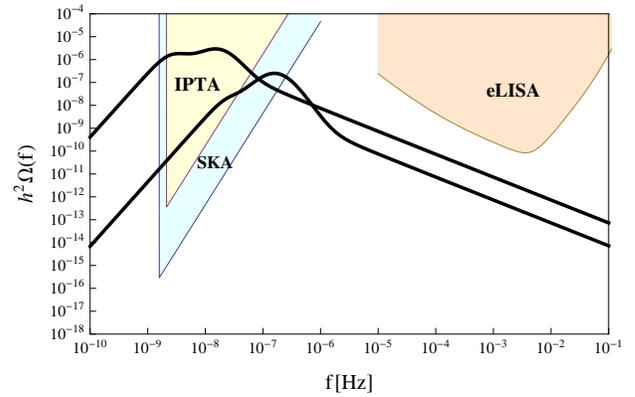}\caption{\label{f1} The spectrum of GWs from a QCD PT (for $ N=3 $) along with the sensitivity of three detectors \cite{Moore:2014lga} are displayed. The upper curve is plotted for $ \tau =H_* $ and the lower curve with $ \tau =10H_* $. For IPTA and SKA detectors these GWs would be detectable while eLISA cannot detect their signals. The sensitivity of detectors is based on \cite{site} 20  pulsars with $ 10^{-7}\mathrm{sec} $ timing precision in 15-year observation time for IPTA, and 100 pulsars with $ 3\times  10^{-8}\mathrm{sec} $ timing precision in 20-year observation time for SKA.}
\end{figure}
As seen from FIG. \ref{f1}, described GWs are detectable by IPTA.\\ IPTA is a project consisted of European Pulsar Timing Array (EPTA), Parkes Pulsar Timing Array (PPTA), and North American Nanohertz Observatory for Gravitational Waves (NANOGrav). They use millisecond pulsars to detect passing GW sensitive to the frequency range $ 10^{-10}~\mathrm{Hz}-10^{-7}~\mathrm{Hz} $. (Pulsars are rotating neutron stars which radiate electromagnetic waves and can signal a passing GW causing a fluctuation in the arrival time of their pulses.)\\The other relevant detector is SKA which is a huge radio telescope and can detect GWs by using pulsars in the next decade. However, eLISA \cite{elisa} space-based interferometer scheduled in 2034 will not be able to detect these GWs.
\subsection{Soft Wall}
\begin{figure}[th]
\includegraphics[scale=.92]{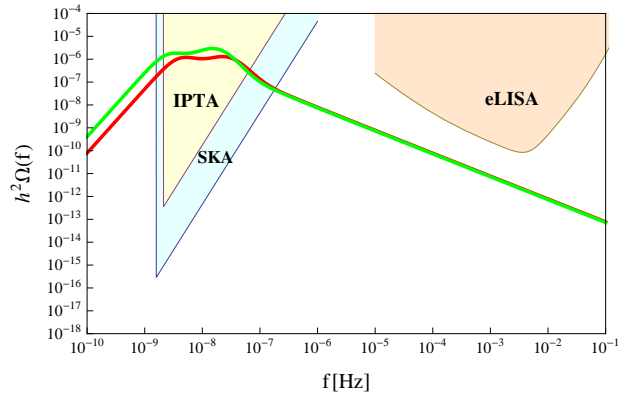}\caption{\label{f3} Calculation results of hard and soft wall models are compared for $ N=3 $ and $ \tau =H_* $. The upper curve shows the GW estimated in the hard wall model.}
\end{figure}

\begin{figure}[th]
\includegraphics[scale=.92]{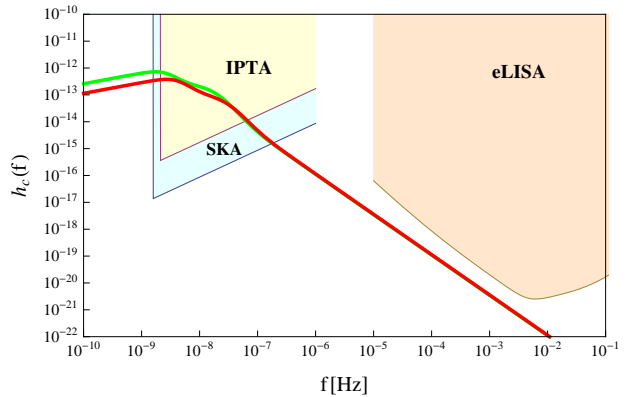}\caption{\label{f4}The characteristic strain amplitude of the GWs are plotted for $ N=3 $ and $ \tau =H_* $ in the hard and soft wall models. The upper curve is estimated in hard wall model.}
\end{figure}

In this case, the gravity action is
\begin{equation}\label{22}
S=-\frac{1}{16\pi G_5}\int d^5x \sqrt{g}~e^{-\phi}\Big(\mathcal{R}+\frac{12}{R^2}\Big),
\end{equation}
where the dilaton field is denoted as $ \phi=cz^2 $ which in fact is a smooth cap off. Assuming the dilaton field does not backreact on the metric, AdS and AdS-BH are solutions of the e.o.m.\\Similar to the previous section calculation, one can acquire the free energy density values by using the same conditions mentioned in the hard wall model
\begin{eqnarray}\label{23}
F^{AdS}&\simeq&\frac{4R^3T}{8\pi G_5}\int _0^{\beta '} dt\int _{\varepsilon}^{\infty}dz~z^{-5}e^{-cz^2}\nonumber \\&=&\frac{R^3}{8\pi G_5}\Big (c^2(\frac{3}{2}-\gamma)+\frac{1}{\varepsilon ^4}-\frac{2c}{\varepsilon ^2}-c^2\ln (c\varepsilon ^2)-\frac{1}{2z_h^4}\Big), \nonumber \\
\end{eqnarray}
\begin{eqnarray}\label{24}
F^{BH}&\simeq &\frac{4R^3T}{8\pi G_5}\int _0^{\pi z_h} dt\int _{\varepsilon}^{z_h}dz~z^{-5}e^{-cz^2}\nonumber\\ &=&\frac{R^3}{8\pi G_5}\Big (c^2(\frac{3}{2}-\gamma)+c^2\mathrm{Ei}(-cz_h^2)+e^{-cz_h^2}(\frac{c}{z_h^2}-\frac{1}{z_h^4}) \nonumber \\&+&\frac{1}{\varepsilon ^4}-\frac{2c}{\varepsilon ^2}-c^2\ln (c\varepsilon ^2)\Big)
\end{eqnarray}
where $ \mathrm{Ei}(x)\equiv -\int _{-x}^{\infty}dt~e^{-t}/t  $ and $ \gamma \sim 0.5 $. By calculating $ \Delta F $, one finds there exists a PT for $ cz_h^2=0.419 $ and thus $ T_*=0.492 \sqrt{c} $. From calculations of the lightest $ \rho $ meson mass in the soft model \cite{Karch:2006pv}, $ \sqrt{c} =388~\mathrm{MeV} $. This leads to $ T_*=191~\mathrm{MeV} $.\\ Also, we can calculate $ \alpha $ from $ \Delta F $ and by the same argument in the hard wall model, it is found that for a string length of the order of $ l_s\sim\sqrt{\alpha _s} \sim 1/(10^{9}~\mathrm{MeV})-1/(10^{8}~\mathrm{MeV}) $, $ \tau =H_* $ and $ \tau =10H_* $, respectively. Different values of the soft wall model compared to the hard wall model stem from different IR cut-off in the hard wall, $ z_0 $, and the soft wall, $ c $, actually in $ \phi =c z^2 $. Therefore, this leads to different transition temperature, and consequent latent heat. Here, again the QCD PT imprint on the GWs is traceable by IPTA and SKA detectors, FIG. \ref{2}.\\ In FIG. \ref{f3} and \ref{f4}, the comparison of the GW estimation in the hard and soft wall models is shown. The spectral shape of the GWs is differently scaled with respect to small and large frequencies (Eq. (\ref{3})). For $ \tau =H_* $, due to the dominant energy density contribution of bubbles, envelope approximation, this would be almost $ 10^{19} f^3 $ and $ 10^{18} f^3 $ with small frequencies for the hard and soft wall model, respectively, and for larger frequencies approximately $ 10^{-14} f^{-1} $ for both models. Also, as a result of different $ \alpha $, GW spectrum of these two models becomes more distinctive for small frequencies.
\begin{figure}[th]
\includegraphics[scale=0.92]{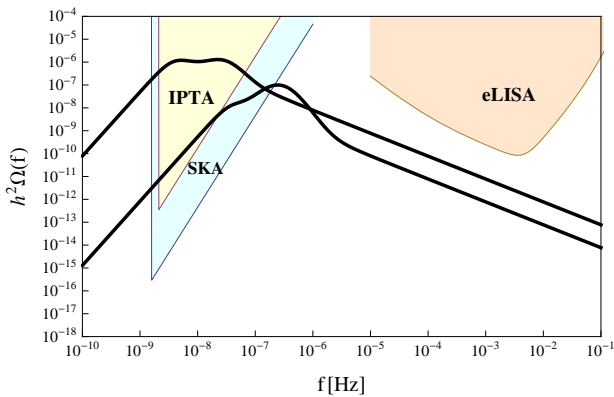}\caption{\label{f2} For the soft wall model the GW spectrum is plotted, with the same conditions mentioned in FIG. \ref{f1}.}
\end{figure}
\\In summary, during the evolution of the universe, for heavy non-dynamical quarks confinement-deconfinement PT was a first order type with non-zero and zero expectation value of Polyakov loop, as the order parameter, in the deconfined and confined phase, respectively. We studied this PT by AdS/QCD approach through GWs produced by the bubble collision, sound waves and MHD turbulence.\\ In the hard and soft wall models of QCD, there exists a first order Hawking-Page PT corresponded to the confinement-deconfinement PT. Using this correspondence, we estimated the GW spectrum associated with these sources during the QCD PT in these models through numerical simulations.\\We assumed that the duration of the PT corresponds to the BH evaporation time in the five dimensional AdS space. Hence, $ \tau =H_* $, $\tau ^{-1} \simeq 10^{-5}~ \mathrm{sec} $, as the scale of PT duration leads to $ l_s \sim  1/(10^{10}~\mathrm{MeV})-1/(10^{9}~\mathrm{MeV}) $ for the string length in the hard and soft wall models, respectively. We also calculated the latent heat at the transition and obtained radiated GWs for $ N=3 $ in the models. The peak frequency of GWs, which IPTA and SKA will be able to detect their signals as an evidence for this PT, falls in $ 10^{-8}~\mathrm{Hz}-10^{-7}~\mathrm{Hz} $ band.
\begin{acknowledgments} \label{Calculation}
We thank M. M. Sheikh-Jabbari for useful comments.
\end{acknowledgments}

 \end{document}